\documentclass[aps,pra,preprint,superscriptaddress,longbibliography]{revtex4-2}

\usepackage{amsmath}
\usepackage{amsthm}
\usepackage{amsfonts}
\usepackage{amssymb}
\usepackage{xcolor,graphicx}
\usepackage{tikz}
\usepackage{color}
\usepackage[pdftex]{hyperref} 
\usepackage{longtable}
\usepackage{array}
\usepackage{dsfont}

\begin{document}
	
	\title{Quadratic Time-Dependent Quantum Harmonic Oscillator}
	
	\author{F. E. Onah}
	\email[e-mail: ]{A00834081@tec.mx}
	\affiliation{Tecnologico de Monterrey, Escuela de Ingenier\'ia y Ciencias, Ave. Eugenio Garza Sada 2501, Monterrey, N.L., Mexico, 64849}
    \affiliation{The Division of Theoretical Physics, Physics and Astronomy, University of Nigeria Nsukka, Nsukka Campus, Enugu State, Nigeria.}
		
	\author{E. Garc\'ia Herrera}
	\email[e-mail: ]{A01274795@tec.mx}
	\affiliation{Tecnologico de Monterrey, Escuela de Ingenier\'ia y Ciencias, Ave. Eugenio Garza Sada 2501, Monterrey, N.L., Mexico, 64849}	

	\author{J. A. Ruelas-Galv\'an}
	\email[e-mail: ]{A01424274@tec.mx}
	\affiliation{Tecnologico de Monterrey, Escuela de Ingenier\'ia y Ciencias, Ave. Eugenio Garza Sada 2501, Monterrey, N.L., Mexico, 64849}	

	\author{G. Ju\'arez Rangel}
	\email[e-mail: ]{A01570628@tec.mx}
	\affiliation{Tecnologico de Monterrey, Escuela de Ingenier\'ia y Ciencias, Ave. Eugenio Garza Sada 2501, Monterrey, N.L., Mexico, 64849}	

	\author{E. Real Norzagaray}
	\email[e-mail: ]{A01731296@tec.mx}
	\affiliation{Tecnologico de Monterrey, Escuela de Ingenier\'ia y Ciencias, Ave. Eugenio Garza Sada 2501, Monterrey, N.L., Mexico, 64849} 	
	
	\author{B. M. Rodr\'iguez-Lara}
    \email[e-mail: ]{bmlara@tec.mx}
	\affiliation{Tecnologico de Monterrey, Escuela de Ingenier\'ia y Ciencias, Ave. Eugenio Garza Sada 2501, Monterrey, N.L., Mexico, 64849}	
	
	\date{\today}
	
\begin{abstract}
    We present a Lie algebraic approach to a Hamiltonian class covering driven, parametric quantum harmonic oscillators where the parameter set---mass, frequency, driving strength, and parametric pumping---is time-dependent. 
    Our unitary-transformation-based approach provides a solution to our general quadratic time-dependent quantum harmonic model. 
    As an example, we show an analytic solution to the periodically driven quantum harmonic oscillator without the rotating wave approximation; it works for any given detuning and coupling strength regime. 
    For the sake of validation, we provide an analytic solution to the historical Caldirola--Kanai quantum harmonic oscillator and show that there exists a unitary transformation within our framework that takes a generalized version of it onto the Paul trap Hamiltonian.
    In addition, we show how our approach provides the dynamics of generalized models whose Schr\"odinger equation becomes numerically unstable in the laboratory frame.   
\end{abstract}
	
	
	\maketitle
	\newpage

\section{Introduction}

The quantum harmonic oscillator (QHO) is an essential physical modeling tool for a wide range of theoretical and experimental problems from electrodynamics \cite{More1970p2679} to cosmology \cite{Polarski1996p377}.
Since the quantization of the classical harmonic oscillator for time-independent parameters \cite{Born1924p379, Heisenberg1925p879, Born1926p557, Heisenberg1927p172}, the QHO evolved to meet the needs of the time.
The push to describe non-conservative systems \cite{Bateman1931p815} lead to the Caldirola-Kanai model \cite{Caldirola1941p393, Kanai1948p440}, where the mass is time-dependent and decays exponentially, that, in time, proved to arise from parametric processes and not dissipation \cite{SPKim1994p3927}. Nevertheless, it served as a standard model for trying to introduce dissipation \cite{Dekker1981p0370, Razavy2005p576, Cordero2010p1884, Kochan2010p022112, Deguchi2020p022105}.
After the Caldirola-Kanai model, describing the motion of a charged particle motivated the study of the QHO with time-dependent frequency, like quantum motion in a Paul trap \cite{Brown1991p527}, that led to the Lewis-Riesenfield-Ermakov invariant \cite{Lewis1968p1976, LewisRiesenfield1967p510, LewisRiesenfield1969p1458}, and models with both time-dependent mass and frequency \cite{Khandekar1975p384, Yeon1994p1035, SPKim1999p2711, Pedrosa1997p3129, SPKim1995p4268, LewisRiesenfield1969p1458, Kanasugi1995p949}.
Time-dependent forcing in the position lead to the driven QHO \cite{Dodonov1979p550,CFLo1993p115,Choi2003p321,Mandal2003p983,Rahav2003p013820,Buyukasik2018p082104,Santana2019p1900035,Guedes2001p034102} where the first solution involved a Gaussian ansatz \cite{Husimi1953p381}.

We are interested in a general quadratic time-dependent QHO \cite{Maamache1996p833,Maamache1998p161,Song2002p0,Yeon2003p052108,Singh2010p4685, IbarraSierra2015p83, Santana2016p042104}, 
\begin{equation}\label{eq:GTDQHO}
    \hat{H}(t) = \frac{1}{2 m(t)} \hat {p}^{2} + \frac{1}{2} m(t) \omega^{2}(t) \hat{x}^{2} + \alpha_{x}(t) \hat{x} + \alpha_{p}(t) \hat{p} + \alpha_{xp}(t) \left\{ \hat{x}, \hat{p} \right\} + \alpha_{0}(t),
\end{equation}
including driving to the position and momentum, third and fourth terms in the right-hand-side in that order, as well as parametric processes, fifth right-hand-side term. 
It is possible to solve a simpler version of this Hamiltonian, without the parametric processes, using different approaches like the Feynman Path integral \cite{Yeon1987p611, Cohen1998p537, Song2002p0, Dorofeyev2016p200}, generalizations of the Lewis-Riesenfield-Ermakov invariant  \cite{Kumar2021p112705, Bekkar2003p016101}, or unitary transformations that diagonalize the model \cite{Maamache1996p833, Maamache1998p161, Santana2016p042104} with the help of algebraic Lie methods \cite{Boldt2013p022116, Lo1997p144, Lo1993p319}.
Understanding the model may help optimizing plasmon lifetimes in nanostructures \cite{Yildiz2020p035416}, bring forward new schemes for quantum machines \cite{Bennett2020p103028} and heat engines \cite{Chen2020p124140} to explore quantum thermodynamics \cite{Aoki2021p052208, Hyeong2022p127974}, improve quantum data transmission and processing  \cite{Deimert2020p097403, Castanos2021p033709, Qasymeh2022p0, Lorenzo2022p023030, Liu2021p032605}, bring forward new approaches to condensed matter \cite{Braganca2020p0}, quantum field theory \cite{Verge2009p1360}, or even gravitational interactions \cite{Bose2022p106028, Chakraborty2022p115691}.

Here, we aim to diagonalize our full general quadratic time-dependent QHO using algebraic Lie methods \cite{Wei1963p575,Rau1998p4785} in first quantization.
In the Results Section, we present the time-independent model and its symmetries. 
We use these symmetries to diagonalize the time-independent model and establish a road map for the diagonalization of the time-dependent model.
Then, we diagonalize the time-dependent model and calculate the expectation value of its observables and its variances. 
We analytically solve the time-dependent harmonically driven QHO without the rotating wave approximation, the Caldirola-Kanai model, and a quadratic QHO with random time-dependent parameter functions in the Discussion Section in order to exemplify and discuss our method.
We close with our conclusions.
 
\section{Time-independent model} \label{sec:TimeIndependentModel}
We are interested in a family of QHOs including all possible position and momentum combinations up to second order \cite{Lo1993p319},
\begin{align}\label{eq:GTIQHO}
    \hat{H} = \frac{ 1 }{ 2m } \hat{p}^{2} + \frac{ 1 }{ 2 } m \omega^{2} \hat{x}^{2} + \alpha_{x} \hat{x} + \alpha_{p} \hat{p} + \alpha_{xp} \{ \hat{x} , \hat{p} \} + \alpha_{0}, 
\end{align}
where the mass is given by $m$, the oscillator frequency by $\omega$, the external driving of the position by $\alpha_{x}$ with units of $\mathrm{energy} \cdot \mathrm{position}^{-1}$, the constant shift of the momentum by $\alpha_{p}$ with units of $\mathrm{energy} \cdot \mathrm{momentum}^{-1}$, the external scaling constant or parametric drive $\alpha_{xp}$ with units of $\mathrm{time}^{-1}$ and a constant energy $\alpha_{0}$.
We use the standard notation for the anti-commutator $\{\hat{x},\hat{p}\} = \hat{x} \hat{p} + \hat{p} \hat{x}$.

The Hamiltonian model shows a closed underlying algebra,
\begin{equation}
\begin{aligned}
    \left[\hat{x},\hat{p}\right]&=i\hbar,&\left[\{\hat{x},\hat{p}\},\hat{x}\right]&=-2i\hbar\hat{x},\\
    \left[\hat{x},\hat{p}^{2}\right]&=2i\hbar\hat{p},&\left[\{\hat{x},\hat{p}\},\hat{p}\right]&=2i\hbar\hat{p},\\
    \left[\hat{p},\hat{x}^{2}\right]&=-2i\hbar\hat{x},&\left[\{\hat{x},\hat{p}\},\hat{x}^{2}\right]&=-4i\hbar\hat{x}^{2}, \\
    \left[\hat{x}^{2},\hat{p}^{2}\right]&=2i\hbar\{\hat{x},\hat{p}\},&\left[\{\hat{x},\hat{p}\},\hat{p}^{2}\right]&=4i\hbar\hat{p}^{2}, \\
\end{aligned}  
\end{equation}
that help us define a group with three types of unitary transformations. First, a translation,
\begin{equation}
        \begin{aligned}
            e^{ \frac{i}{\hbar} \left[ \beta_{p} \hat{x} + \beta_{x} \hat{p}  \right] } \, \hat{x} \,  e^{ - \frac{i}{\hbar} \left[  \beta_{p} \hat{x} + \beta_{x} \hat{p} \right]} & = \hat{x} + \beta_{x}, \\
            e^{ \frac{i}{\hbar} \left[ \beta_{p} \hat{x} + \beta_{x} \hat{p} \right] } \, \hat{p} \,  e^{ - \frac{i}{\hbar}  \left[ \beta_{p} \hat{x} + \beta_{x} \hat{p} \hat{x}\right]} &= \hat{p} - \beta_{p},
        \end{aligned}
\end{equation}
where the real constant displacement parameters $\beta_{x}$ and $\beta_{p}$ have units of $\mathrm{position}$ and $\mathrm{momentum}$, respectively. 
A rotation, 
\begin{equation}
\begin{aligned}
    e^{\frac{i}{\hbar} \left[ \theta_{x}^{2} \hat{x}^{2} + \theta_{p}^{2} \hat{p}^{2} \right]} \, \hat{x} \, e^{-\frac{i}{\hbar}\left[ \theta_{x}^{2} \hat{x}^{2} + \theta_{p}^{2} \hat{p}^{2} \right]} &= \cos{\left(2 \theta_{x} \theta_p \right)} \, \hat{x} + \frac{ \theta_{p}}{\theta_{x} } \sin{\left(2 \theta_{x}  \theta_p \right)} \,  \hat{p}, \\
    e^{\frac{i}{\hbar}\left[ \theta_{x}^{2} \hat{x}^{2} + \theta_{p}^{2} \hat{p}^{2}\right]} \, \hat{p} \, e^{-\frac{i}{\hbar}\left[\theta_{x}^{2} \hat{x}^{2} + \theta_{p}^{2} \hat{p}^{2}\right]} &= \cos{\left(2  \theta_{x}  \theta_p \right)} \, \hat{p} - \frac{\theta_{x}}{ \theta_{p}} \sin{\left(2 \theta_{x}  \theta_p \right)} \,  \hat{x},
    \label{rotation}
\end{aligned}
\end{equation}
where the rotation parameters $\theta_{x}^{2}$ and $\theta_{p}^{2}$ have units of $\mathrm{momentum} \cdot \mathrm{position}^{-1}$ and $\mathrm{position} \cdot \mathrm{momentum}^{-1}$, respectively.
It will be helpful later to note that these rotations take a simpler form, 
\begin{equation}
\begin{aligned}
    \lim_{\theta_{x} \rightarrow 0} e^{\frac{i}{\hbar}\left[\theta_{x}^{2} \hat{x}^{2} + \theta_{p}^{2} \hat{p}^{2}\right]} \, \hat{x} \, e^{-\frac{i}{\hbar}\left[\theta_{x} \hat{x}^{2} + \theta_{p} \hat{p}^{2}\right]} &=  \hat{x} + 2 \theta_{p}^{2} \, \hat{p}, \\
    \lim_{\theta_{p} \rightarrow 0} e^{\frac{i}{\hbar}\left[\theta_{x}^{2} \hat{x}^{2} + \theta_{p}^{2} \hat{p}^{2}\right]} \, \hat{p} \, e^{-\frac{i}{\hbar}\left[\theta_{x}^{2} \hat{x}^{2} + \theta_{p}^{2} \hat{p}^{2}\right]} &= \hat{p} - 2 \theta_{x}^{2} \, \hat{x},
\end{aligned}
\end{equation}
when one of the parameters is zero. 
Finally, we have a squeezing transformation, 
\begin{equation}
        \begin{aligned}
             e^{ \frac { i }{ \hbar } \beta \{\hat{x},\hat{p}\} } 
                 \, \hat{x} \, e^{ - \frac{ i }{ \hbar } \beta \{ \hat{x} , \hat{p} \} } &= \hat{x} e^{2 \beta}, \\
              e^{ \frac{ i }{ \hbar } \beta \{ \hat{x} , \hat{p} \} } 
                 \, \hat{p} \,  e^{ - \frac{ i }{ \hbar } \beta \{ \hat{x} , \hat{p} \} } &= \hat{p} e^{ -2 \beta },
        \end{aligned}
\end{equation} 
that provides inverse scaling for position and momentum with a dimensionless real scaling parameter $\beta$. 

In order to create insight, we use these transformations to diagonalize our Hamiltonian in Eq. (\ref{eq:GTIQHO}).
First, we move into a reference frame,
\begin{align} 
    \vert {\Psi} \rangle = e^{-\frac {i} {\hbar} \alpha_{0} t} \vert {\psi_{1}} \rangle,
\end{align}
rotating at the constant frequency $\alpha_{0}$ to obtain the effective Hamiltonian,
\begin{equation}
    \hat {H_{1}} = \hat {H} - \alpha_{0},
\end{equation}
without the constant bias.
In order to deal with the forced position and momentum terms proportional to $\hat{x}$ and $\hat{p}$, we apply the translation,
\begin{equation}
     \vert {\psi_{1}} \rangle = e^{-\frac {i} {\hbar} \left( \beta_{p} \hat{x} + \beta_{x} \hat{p} \right)} \vert{\psi_{2}} \rangle, 
\end{equation}
with parameters,  
\begin{equation}
\begin{aligned}  \label{BxBp_Independent}
    \beta_{x} &=  \frac { 2 m \alpha_{p} \alpha_{xp} - \alpha_{x} } {m(\omega^{2} - 4 \alpha_{xp}^{2}) } , \\
    \beta_{p} &= m\alpha_{p} - \frac{2\alpha_{xp}(\alpha_{x}-2 m \alpha_{p}\alpha_{xp})}{\omega^2-4\alpha_{xp}^2} ,
\end{aligned}
\end{equation}
that yields an effective Hamiltonian,
 \begin{equation}
     \hat {H_{2}} = \frac {1}{2m} \hat{p}^{2} + \frac{1}{2} m \omega^{2} \hat {x}^{2}  + \alpha_{xp} \{\hat{x},\hat{p}\} + l,
 \end{equation}
where the accumulation constant,
\begin{equation}
    l = \frac {1} {2m}\beta_{p}^{2} + \frac {1} {2} m \omega^{2} \beta_{x}^{2} + \alpha_{x} \beta_{x} - \alpha_{p} \beta_{p} - 2 \alpha_{xp} \beta_{x} \beta_{p},
\end{equation}
defines a new reference frame, 
\begin{equation}
     \vert {\psi_{2}} \rangle = e^{-\frac {i} {\hbar} l t} \vert {\psi_{3}} \rangle,
\end{equation}
with an effective Hamiltonian, 
\begin{equation}
    \hat{H_{3}} =  \frac{1}{2m} \hat{p}^{2} + \frac{1}{2} m\omega^{2}\hat{x}^{2} + \alpha_{xp} \{\hat{x},\hat{p}\},
\end{equation}
that a rotation,
\begin{equation}
    \vert {\psi_{3}} \rangle  = e^{-\frac{i}{\hbar}(\theta_{x}^{2} \hat{p}^{2} + \theta_{p}^{2} \hat{x}^{2})} \vert{\psi_{4}} \rangle,
\end{equation}
with parameters fulfilling the transcendental equation, 
\begin{equation}
      \tan \left( 4 \theta_{x} \theta_{p} \right)  =  \frac{m \alpha_{xp}}{\theta_{x}^{2} - m^{2} \omega^{2} \theta_{p}^{2}} \, 4 \theta_{x} \theta_{p},
\end{equation}
translates into the standard harmonic oscillator, 
\begin{equation}
    \hat{H}_{4} = \frac{1}{2 M} \hat{p}^{2} + \frac{1}{2} M \Omega^{2} \hat{x}^{2},
\end{equation}
with effective mass and frequency,
\begin{equation} \label{M_Omega_TI}
    \begin{aligned} 
        M =& 2m \left[ 1 +   m^{2} \omega^{2} \frac{\theta_{p}^{2}}{\theta_{x}^{2}} + \left( 1 -  m^{2} \omega^{2} \frac{\theta_{p}^{2}}{\theta_{x}^{2}} \right) \cos \left( 4 \theta_{x} \theta_{p}  \right)  + 4 m \alpha_{xp} \frac{\theta_{p}}{\theta_{x}} \sin \left( 4 \theta_{x} \theta_{p}  \right) \right]^{-1}, \\
        \Omega^{2} =& \frac{1}{2 m M} \frac{\theta_{x}^{2}}{\theta_{p}^{2}}\left[1 + m^{2} \omega^{2} \frac{\theta_{p}^{2}}{\theta_{x}^{2}}  - \left( 1 - m^{2} \omega^{2} \frac{\theta_{p}^{2}}{\theta_{x}^{2}} \right) \cos \left( 4 \theta_{x} \theta_{p}  \right) - 4 m \alpha_{xp} \frac{\theta_{p}}{\theta_{x}} \sin \left( 4 \theta_{x} \theta_{p}  \right) \right],
    \end{aligned}
\end{equation}
in terms of the real parameters from our original general quadratic time-independent QHO model.
At this point, it is important to realize that three simple parameter choices are available at hand. 
First, we may choose to perform the rotation with just the position term, $\theta_{p} = 0$, to obtain, 
\begin{equation}
\begin{aligned}
    \theta_{x}^{2} &= m \alpha_{xp} , \\
    \lim_{\theta_{p} \rightarrow 0} M &= m  , \\
    \lim_{\theta_{p} \rightarrow 0} \Omega^2 &=  \omega^2 - 4 \alpha_{xp}^{2} , \\
\end{aligned}
\end{equation}
a harmonic oscillator with effective mass identical to that in the original frame and effective frequency that tells us the dimensional external scaling should not overcome a fourth of the resonant frequency, $\omega^{2} > 4 \alpha_{xp}^2$, in order to have a proper harmonic oscillator in the diagonal frame. 
We may choose to perform the rotation with just the momentum term, $\theta_{x} = 0$, 
\begin{equation}
\begin{aligned}
    \theta_{p}^{2} &=  -\frac{\alpha_{xp}}{m \omega^{2}} , \\
    \lim_{\theta_{x} \rightarrow 0} M &= \frac{m \omega^2}{\omega^2 - 4 \alpha_{xp}^{2}} , \\
    \lim_{\theta_{x} \rightarrow 0} \Omega^2 &= \omega^2 - 4 \alpha_{xp}^{2} , \\
\end{aligned}
\end{equation}
and recover an effective mass scaled by the effective frequency square, that is identical to the later case. 
We may take the parameter values equal in magnitude, $\theta_{x} = \theta_{p} = \theta$, and realize that the effective frequency remains the same but the effective mass becomes more complicated in shape. 
We feel it is important to stress that the absolute value of the dimensional external scaling $\alpha_{xp}$ must be restricted to half the value of the free oscillator frequency, $ \vert \alpha_{xp} \vert < \omega / 2 $, in order deal with a harmonic oscillator and avoid dealing with a free particle or an oscillator with complex effective frequency.

\section{Time-dependent model}\label{sec:TimeDependentModel}

Now, let us focus on our general quadratic time-dependent QHO Hamiltonian, Eq. (\ref{eq:GTDQHO}),
\begin{equation}\label{eq:GTDQHOH}
    \hat{H}(t) = \frac{1}{2 m(t)} \hat {p}^{2} + \frac{1}{2} m(t) \omega^{2}(t) \hat{x}^{2} + \alpha_{x}(t) \hat{x} + \alpha_{p}(t) \hat{p} + \alpha_{xp}(t) \left\{ \hat{x}, \hat{p} \right\} + \alpha_{0}(t), \nonumber
\end{equation}
where the units of all the coefficients are the same as in the time-independent Hamiltonian but they are now well-behaved functions of time.

In order to diagonalize our Hamiltonian, we follow a road map similar to the time-independent diagonalization.
Our first transformation, again, moves into a rotating frame, 
\begin{equation}\label{eq:FirstTDQHO}
    \vert {\Psi} \rangle = e^{-\frac {i} {\hbar} \int_0^t \alpha_{0} \left( \tau \right) d \tau} \vert {\psi_{1} } \rangle,
\end{equation}
yielding an effective Hamiltonian,
\begin{equation}
    \hat {H}_1(t) = \frac {1} {2 m(t)} \hat {p}^{2} + \frac {1} {2} m(t) \omega^{2}(t) \hat {x}^{2} + \alpha_{x} (t) \hat {x} + \alpha_{p} (t) \hat{p} + \alpha_{xp} (t) \{ \hat{x} , \hat{p} \},
\end{equation}
without the time-dependent energy bias term.
Now, a time-dependent translation, 
\begin{equation} \label{Eliminate_x_p_terms}
    \vert {\psi_{1}} \rangle = e^{- \frac {i} {\hbar} \left[ \beta_{p}(t) \hat {x} + \beta_{x}(t) \hat {p} \right] } \vert {\psi_2} \rangle,
\end{equation}
with parameters solving the differential equation set, 
\begin{equation}
\begin{aligned}
    \ddot {\beta}_{x}(t) & + \frac {\dot{m} (t)} {m (t)} \dot {\beta}_{x} (t) - \left[ 2 \dot {\alpha}_{xp} (t) - \omega^{2} (t) + 4 \alpha_{xp}^{2} (t) + \frac {2 \alpha_{xp} \dot {m} (t)} {m (t)} \right] \beta_{x} (t) \ldots\\
    &=\dot{\alpha}_{p}(t) - \frac {\alpha_{x}(t)} {m(t)} + 2 \alpha_{p}(t) \alpha_{xp}(t) +
    \alpha_{p}(t) \frac {\dot{m}(t)} {m(t)},\\
    \beta_{p}(t) &= m (t) \left[ - \dot {\beta}_{x}(t) + 2 \alpha_{xp}(t) \beta_{x}(t) + \alpha_{p}(t) \right],
\end{aligned}
\end{equation}
where the boundary conditions are given by Eq.(\ref{BxBp_Independent}) in the time-independent case. It helps us obtain an effective Hamiltonian without the linear driving of the configuration variables, 
\begin{equation}\label{eq:STTDHO}
    \hat {H}_{2} (t) = \frac {1} {2 m (t)} \hat {p}^{2} + \frac {1} {2} m (t) \omega^{2} (t) \hat {x}^{2} + \alpha_{xp} (t) \{ \hat {x}, \hat {p} \} + \ell (t),
\end{equation}
where the accumulation time-dependent energy bias term,
\begin{equation}
    \ell (t) =  \frac {1} {2 m (t)} \beta_{p}^{2} (t) + \frac {1} {2} m (t) \omega^{2} (t) \beta_{x}^{2} (t) + \alpha_x (t) \beta_{x} (t) - \alpha_{p} (t) \beta_{p} (t) - 2 \alpha_{xp} (t) \beta_{x} (t) \beta_{p} (t),
\end{equation}
commutes with all other terms, and help's define a rotating frame,
\begin{equation}
    \vert {\psi_{2}} \rangle = e^{ - \frac{ i }{ \hbar } \int_0^{t} \ell( \tau ) d \tau } \vert{\psi_{3}} \rangle,
\end{equation}
which yields the effective Hamiltonian,
\begin{equation}\label{eq:QHOSqueezedFrame}
    \hat{H}_{3} \left( t \right ) = \frac{ 1 }{ 2 m (t) } \hat{p}^{2} + \frac{ 1 }{ 2 } m(t) 
    \omega^{2} (t) \hat{x}^{2} + \alpha_{xp} (t) \{ \hat{x} , \hat{p} \},
\end{equation}
that is the well-known time-dependent QHO with an extra driving in the anti-commutation term; that is, a driven parametric QHO.
Here, we need to take a slight deviation from the previous road map and use the squeezing transformation,
\begin{equation}
    \vert {\psi_{3}} \rangle =  e^{- \frac{i}{\hbar} \frac{\ln \left[ \gamma \left( t \right) \right]} {2} \left\{\hat{x} , \hat{p} \right\}} \vert {\psi_{4}}\rangle,
\end{equation}
where the new dimensionless parameter, 
\begin{equation}
    \gamma^{2} (t) = \frac{m(0)\omega(0)}{m (t) \omega (t)},
\end{equation}
given in terms of the time-dependent mass and frequency of the system as well as their initial time values, yields the effective parametric QHO Hamiltonian,
\begin{equation} \label{eq:parametric_QHO_Hamiltonian}
\hat{H}_4 (t) = \frac{\omega (t)}{\omega(0)} \left[ \frac{1}{2 m(0) } \hat{p}^{2} + \frac{1}{2} m(0) \omega^{2}(0)  \hat{x}^2 \right] +  \frac{1}{2}\kappa(t) \left\{\hat{x}, \hat{p}\right\},
\end{equation}
where we define an effective parametric drive,
\begin{equation} \label{eq:kappa}
\kappa (t) = \frac{1}{2}\left[ \frac{\dot{m}(t)}{m(t)} + \frac{\dot{\omega}(t)}{\omega(t)} + 4\alpha_{xp}(t) \right],
\end{equation}
for the sake of space. 
The first right-hand-side of this effective Hamiltonian is the factor of a time-dependent term and a time-independent QHO term suggesting a rotation,
\begin{equation}\label{eq:rotation_transformation}
    \vert {\psi_4} \rangle = e^{- \frac{i}{\hbar\omega(0)} \left[\frac{\hat{p}^{2}}{2 m(0)} + \frac{1}{2} m(0) \omega^{2}(0) \hat{x}^2 \right] \frac{\pi}{4}} \vert {\psi_{5}} \rangle,
\end{equation}
yielding the standard time-dependent QHO, 
\begin{equation}
\hat{H}_{5} (t) = \frac{1}{2 m_{5}(t)} \hat{p}^{2} + \frac{1}{2} m_{5}(t) \omega_{5}^{2}(t)  \hat{x}^2,
\end{equation}
with effective mass and frequency,
\begin{equation}\label{eq:effective_m5_omega5}
    \begin{aligned}
        m_{5}(t)&= \frac{m(0)\omega(0)}{\omega(t) + \kappa(t) }, \\
        \omega_{5}^{2}(t) &= \omega^{2}(t) - \kappa^2(t) ,
    \end{aligned}
\end{equation}
where we must be careful to work with well-behaved functions that provide us with positive effective parameters, $m_{5}(t) >0$ and $\omega_{5}^{2}(t) > 0$.
Now, we may factorize the time-dependence using a transformation composed by a squeezing and a rotation,  \cite{BMRodriguez2014p2083}, 
\begin{equation} \label{eq:Rotation-squeezing transformation}
    \vert {\psi_{5}} \rangle = e^{\frac {i} {\hbar} \frac {m_{5} (t) \dot {\rho} (t)}{2 \rho (t)} \hat {x}^{2}} e^{-\frac {i} {\hbar} \frac {\ln \left[\frac{\rho \left(t \right)}{\rho (0)} \right]}{2}\{\hat{x},\hat{p}\}}  \vert{\psi_{6}} \rangle,
\end{equation}
in that order, where we introduce the auxiliary function $\rho\left(t\right)$ fulfilling the Ermakov equation, 
\begin{equation}
    \ddot {\rho} (t) + \frac {\dot{m}_{5} (t)}{m_{5} (t)} \dot {\rho} (t) + \omega_{5}^{2} (t) \rho (t)   = \frac {1} {m_{5}^{2} (t) \rho^{3} (t)}.
\end{equation}
with boundary conditions,
\begin{equation}
    \begin{aligned}
        \rho(0) &= \frac{1}{\sqrt{m_{5} (0)\omega_{5} (0)}}, \\
        \dot{\rho}(0) &= 0.
    \end{aligned}
\end{equation}
that provides us with a diagonal Hamiltonian, 
\begin{equation}\label{Eq:Diagonal Hamiltonian}
        \hat{H}_{6}\left(t\right) = \frac {1} {\omega_{5} (0) m_{5} (t) \rho^{2} (t)}\left[ \frac {1} {2 m_{5} (0)} \hat {p}^{2} + \frac {1} {2} m_{5} (0) \omega_{5}^{2} (0) \hat {x}^{2} \right],
\end{equation}
where the time dependence is factorized from the operators. 
The second factor in the right-hand-side is a time-independent QHO that yields an effective Hamiltonian, 
\begin{equation}
    \hat{H}_{6}(t) = \hbar \Omega(t) \left( \hat{a}^{\dagger} \hat{a} + \frac{1}{2} \right),
\end{equation}
in second quantization with effective time-dependent frequency, 
\begin{equation}
    \Omega(t) = \frac{\omega(t) + \kappa(t)}{m(0)\omega(0)\rho^{2}(t)},
\end{equation}
with creation and annihilation operators, 
\begin{equation}
    \begin{aligned}
    \hat{a} &= \frac{\rho(0)}{\sqrt{2 \hbar}}\left( \frac{1}{\rho^2(0)}\hat{x} + i\hat{p} \right), \\
    \hat{a}^\dagger &=  \frac{\rho(0)}{\sqrt{2 \hbar}}\left(  \frac{1}{\rho^2(0)}\hat{x} - i\hat{p} \right),
    \end{aligned}
\end{equation}
provided by the initial value of the Ermakov parameter with units of $\mathrm{mass}^{-1/2} \cdot \mathrm{frequency}^{-1/2}$.
We want to stress that we use this second quantization with effective time-independent mass and frequency in the diagonal frame just to recover an expression for the time-dependent effective frequency. 
This mathematical gimmick will not be used for the calculation of expectation values or their variances that will be calculated in first quantization.

At this point, it is straightforward to write the time evolution of the initial state of the system in the original frame,
\begin{align}
    \vert {\Psi (t)} \rangle &= \hat{U}_{1} (t) \hat{U}_{2} (t) \hat{U}_{3} (t) \hat{U}_{4} (t) \hat{U}_{5} (t) \hat{U}_{6} (t) \hat{U}_{3}^{\dagger} (0) \hat{U}_{1}^{\dagger} \left( 0 \right)
    \vert{\Psi} \left( 0 \right) \rangle,
\end{align}
in terms of six unitary transformations, 
\begin{equation}
\begin{aligned}
    \hat{U}_{1} (t) &= e^{- \frac {i} {\hbar} \left[ \beta_{x} (t) \hat {p} + \beta_{p} (t) \hat {x} \right] },\\
    \hat{U}_{2} (t) &= e^{-\frac{i}{\hbar} \ln{\left[\sqrt{\frac{\rho (t)}{\rho(0)} \gamma (t)}\right]} \{\hat{x}, \hat{p}\}}, \\
    \hat{U}_{3} (t) &= e^{-\frac{i}{\hbar} \left[ \frac{1}{2 m(0) \omega(0) m_{5} (0) \omega_{5} (0) \rho^{2} (t)}\hat{p}^{2} + \frac{1}{2} m(0) \omega(0) m_{5} (0) \omega_{5} (0) \rho^{2} (t) \hat{x}^{2} \right]\frac{\pi}{4}}, \\
    \hat{U}_{4} (t) &= e^{\frac{i}{\hbar} \frac{m_{5} (0) \omega_{5} (0)}{2}\dot{\rho}(t) \rho (t) m_{5} (t) \hat{x}^{2}},\\
    \hat{U}_{5} (t) &= e^{-\frac {i} {\hbar} \int_0^t \{ \alpha_{0} \left( \tau \right) + \ell \left( \tau \right) \} d \tau}, \\ 
    \hat{U}_{6} \left(t \right) &= e^{-\frac{i}{\hbar}\left[\frac{1}{2m_{5} (0) \omega_{5} (0)}\hat{p}^{2} + \frac{1}{2}m_{5} (0) \omega_{5} (0) \hat{x}^{2} \right]\int_0^t \Omega(\tau) d\tau},
\end{aligned}
\end{equation}
where we group together and simplify some of the transformations defined above, that helps us calculate the expectation values for both position and momentum, 
\begin{equation}\label{eq:Expectation_values}
\begin{aligned}
    \langle \hat{x}(t) \rangle =& A(t) \left[\langle \hat{x}(0) \rangle - \beta_{x}(0)\right] + B(t) \left[\langle \hat{p}(0) \rangle + \beta_{p}(0)\right]+ \beta_{x} (t), \\
    \langle \hat{p}(t) \rangle =& D(t) \left[\langle \hat{x}(0) \rangle - \beta_{x}(0)\right] + E(t) \left[\langle \hat{p}(0) \rangle + \beta_{p}(0)\right]- \beta_{p} (t),
\end{aligned}
\end{equation}
where we use the notation $\langle \hat{o}(\tau) \rangle = \langle \psi(\tau) \vert \hat{o} \vert \psi(\tau) \rangle$ for the expectation values and define the following auxiliary functions, 
\begin{equation}\label{eq:auxiliary_functions}
\begin{aligned}
    A(t) &= \frac{\xi (t)}{2} \left\{\left[ \eta \rho^{2}(0) + \frac{\varepsilon(t)\rho^{2}(0)}{\rho^{2}(t)} - \frac{1}{\eta\rho^{2}(t)} \right] S(t) + \left[ 1 + \frac{\varepsilon(t)}{\eta\rho^{2}(t)} + \frac{\rho^{2}(0)}{\rho^{2}(t)} \right] C(t) \right\},\\
    B(t) &= \frac{\xi (t)}{2\eta} \left\{\left[ \eta\rho^{2}(0) + \frac{\varepsilon(t)\rho^{2}(0)}{\rho^{2}(t)} + \frac{1}{\eta\rho^{2}(t)} \right] S(t) - \left[1 + \frac{\varepsilon(t)}{\eta\rho^{2}(t)} - \frac{\rho^{2}(0)}{\rho^{2}(t)} \right] C(t) \right\},\\
    D(t) &= \frac{\eta}{2\xi(t)}\left\{ \left[ 1 + \frac{\varepsilon(t)}{\eta\rho^{2}(0)} - \frac{\rho^{2}(t)}{\rho^{2}(0)} \right] C(t) - \left[ \eta\rho^{2}(t) - \varepsilon(t) + \frac{1}{\eta\rho^{2}(0)} \right] S(t) \right\}, \\
     E(t) &= \frac{1}{2\xi(t)}\left\{ \left[ 1 - \frac{\varepsilon(t)}{\eta\rho^{2}(0)} + \frac{\rho^{2}(t)}{\rho^{2}(0)} \right] C(t) - \left[ \eta\rho^{2}(t) - \varepsilon(t) - \frac{1}{\eta\rho^{2}(0)} \right] S(t) \right\}, \\
\end{aligned}
\end{equation}
with parameter set,
\begin{equation}
\begin{aligned}
    \eta &= m(0)\omega(0), \\
    \xi (t) &= \frac{\rho(t)}{\rho(0)}\gamma (t), \\
    S (t) &= \sin{\left(\int_{0}^t \Omega \left( \tau \right) d\tau  \right)},  \\
    C (t) &= \cos{\left(\int_{0}^t \Omega \left( \tau \right) d\tau  \right)}, \\
    \varepsilon(t) &= m_{5}(t)\dot{\rho}(t)\rho(t), 
\end{aligned}
\end{equation}
in terms of the functions used to diagonalize the time-dependent Hamiltonian.
The variances for position and momentum, 
\begin{equation}\label{eq:Variances}
\begin{aligned}
    \sigma_{x}^{2}(t) =& A^{2}(t) \sigma_{x}^{2}(0) + B^{2}(t) \sigma_{p}^{2}(0) + A(t)B(t)\left[\langle \left\{\hat{x}(0), \hat{p}(0) \right\} \rangle - 2 \langle \hat{x}(0) \rangle \langle  \hat{p}(0) \rangle \right], \\
    \sigma_{p}^{2}(t) =& D^{2}(t) \sigma_{x}^{2}(0) + E^{2}(t) \sigma_{p}^{2}(0) + D(t)E(t)\left[\langle \left\{\hat{x}(0), \hat{p}(0) \right\} \rangle - 2 \langle \hat{x}(0) \rangle \langle  \hat{p}(0) \rangle \right].
\end{aligned}
\end{equation}
are given in terms of the expected values of position and momentum, $\langle \hat{x}(0) \rangle$ and $\langle \hat{p}(0) \rangle$, and their variances, $\sigma_{x}^{2}(0)$ and $\sigma_{p}^{2}(0)$, for the initial state. 
These expectation values and variances provide a characterization for the time-evolution of initial states where it is straightforward to identify scaling related to the auxiliary function $\xi(t)$ and rotations to $S(t)$ and $C(t)$.

\section{Discussion}\label{sec:Discussion}

In order to provide examples to support our results, we discuss two iconic time-dependent harmonic oscillators, the driven QHO, the Caldirola--Kanai QHO, and a general quadratic QHO with random time-dependent parameters.

\subsection{Driven harmonic oscillator}

In order to provide examples to support our results, we discuss two iconic time-dependent harmonic oscillators, the driven QHO and the Caldirola--Kanai QHO. A driven catity is described by the following Hamiltonian,

\begin{equation} \label{Driven_oscillator}
    \hat{H}{\left(t \right)} = \frac{1}{2m}\hat{p}^{2} + \frac{1}{2}m\omega^{2} \hat{x}^{2} + \Omega \cos{\left(\omega_{d}t\right )} \hat{x},
\end{equation}
where the real constant $\Omega$ is the driving strength of the pump with frequency $\omega_{d}$. 
We obtain this Hamiltonian starting from our general model and considering constant mass, $m(t) = m$, and frequency, $\omega(t) = \omega$, harmonic driving strength, $\alpha_{x}(t) = \Omega \cos (\omega_{d} t)$, of the position, and null values for all other processes, $\alpha_{p}(t) = \alpha_{xp}(t) = \alpha_{0}(t) = 0$.
This system is usually solved under the rotating wave approximation in second quantization and may be feasible of solution following standard Wei-Norman factorization methods used for Floquet engineering \cite{Bandyopadhyay2022p020301}. 
However, it is straightforward to solve it in first quantization using the unitary transformation in Eq.(\ref{Eliminate_x_p_terms}) with time-dependent parameters,
\begin{align}
    \beta_{p}\left(t \right) &= \frac{\Omega \omega_{d} \left[ \omega_{d} \sin{\left( \omega t \right)} - \omega \sin {\left( \omega_{d} t \right) }\right]}{\omega^3 -  \omega \omega_{d}^{2}}, \\
    \beta_{x}\left(t \right) &= \frac{\Omega \omega_{d}^{2} \cos{\left( \omega t \right)} - \omega^{2} \Omega \cos {\left( \omega_{d} t \right) }}{m \omega^4 - m \omega^{2} \omega_{d}^{2}},
\end{align}
arising from the differential equation set, 
\begin{equation}
\begin{aligned}
    \ddot{ \beta }_{ x } (t)  + \omega^{2} \beta_{x} (t) &= - \frac{\Omega}{m} \cos{\left( \omega_{d} t \right )},\\
    \beta_{ p } \left(t \right) &=- m \dot{ \beta }_{x} (t) ,
\end{aligned}
\end{equation}
with initial conditions,
\begin{equation}
\begin{aligned}
    \beta_{x}\left(0 \right) &= -\frac{\Omega}{m \omega^{2}},\\
    \beta_{p} \left(0 \right) &= 0,
\end{aligned}
\end{equation}
leading to the standard harmonic oscillator Hamiltonian, 
\begin{equation}
    \hat{H}_{1} (t) = \frac{1}{2m}\hat{p}^{2} + \frac{1}{2}m\omega^{2} \hat{x}^{2}.
\end{equation}
Since the mass and frequency are time-independent and the external scaling term is zero, the auxiliary functions  from Eq.(\ref{eq:auxiliary_functions}) take the following forms,
\begin{equation}
\begin{aligned}
    A_{d}(t) &= \cos{\left( \omega t \right)}, \\
    B_{d}(t) &= \frac{\sin{\left( \omega t \right)}}{m \omega}, \\
    D_{d}(t) &=  -m\omega \sin{\left( \omega t \right)}, \\
    E_{d}(t) &= \cos{\left( \omega t \right)}, \\
\end{aligned}
\end{equation}
allowing us to obtain the expectation values and variances using the general forms in Eq.(\ref{eq:Expectation_values}) and Eq.(\ref{eq:Variances}), in that order.
It is straightforward to see that the uncertainty principle will depend on the resonant and driving frequencies and the initial variances.

Let us figure out the dynamics of an initial state given by the ground state of the harmonic oscillator,
\begin{align} \label{eq:groundState}
    \psi(0) = \left( \frac{m \omega}{\hbar \pi} \right)^{1/4} \exp \left( - \frac{m \omega x^{2}}{2 \hbar} \right),
\end{align}
with null expectation values at initial time, $\langle \hat{x}(0) \rangle = \langle \hat{p}(0) \rangle = 0$, and variances, $\sigma_{x}^{2}(0) = \hbar / ( 2 m \omega)$ and $\sigma_{p}^{2}(0) = m \omega \hbar/2$, that minimize Heisenberg uncertainty principle, $\sigma_{x}^{2}(0) \sigma_{p}^{2}(0) = \hbar^{\textcolor{blue}{2}}/4$.
The time evolution of these expectation values, 
\begin{align}
    \langle \hat{x}(t) \rangle =&  \frac{\Omega}{m \left(  \omega^{2} - \omega_{d}^{2} \right)} \left[ \cos \left( \omega t \right) - \cos \left( \omega_{d} t \right) \right] , \\
     \langle \hat{p}(t) \rangle =& - \frac{\Omega}{  \omega^{2} - \omega_{d}^{2} } \left[ \omega \sin \left( \omega t \right) - \omega_{d} \sin \left( \omega_{d} t \right) \right], 
\end{align}
provide constant variances, 
\begin{equation}
\begin{aligned}
\sigma_{x}^{2}(t) &= \hbar / ( 2 m \omega), \\ \sigma_{p}^{2}(t) &= m \omega \hbar/2,
\end{aligned}
\end{equation}
that minimize Heisenberg uncertainty principle, $\sigma_{x}^{2}(t) \sigma_{p}^{2}(t) = \hbar^{2}/4$ , as expected from a coherent state.

\begin{figure}[ht]
 	\centering
 	\includegraphics{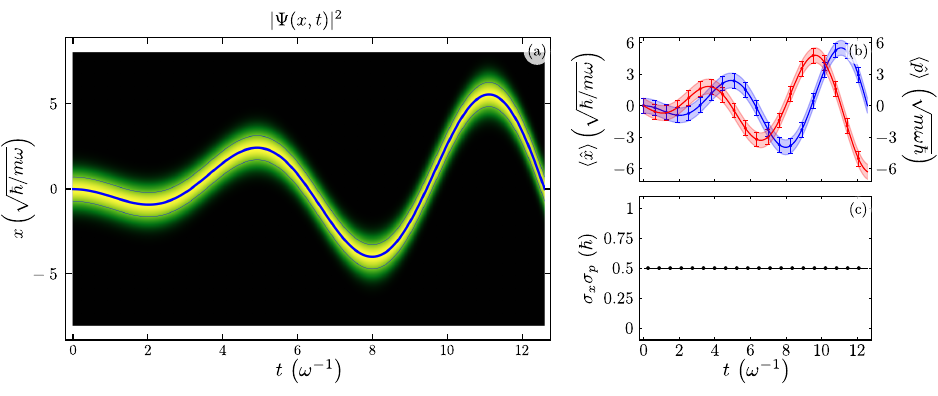}
    \caption{Time evolution of (a) the probability amplitude to find the resonator in the position x $\vert \psi(x) \vert^{2}$, (b) expectation value for the position, $\langle x(t) \rangle$ in blue, and momentum, $\langle p(t) \rangle$ in red, with solid lines and squares for analytic and numeric results, in that order; we show standard deviations, $\sigma_{x}(t)$ and $\sigma_{p}(t)$, as shaded regions or error bars for analytic and numeric results, and (c) Heisenberg uncertainty relation, $\sigma_{x}(t) \sigma_{p}(t)$, for an initial state provided by the ground state of the standard harmonic oscillator for a resonant driven QHO, $\omega_{d} = \omega$. }\label{fig:ResonantDriving}
\end{figure}

Figure \ref{fig:ResonantDriving} shows the evolution of the ground state under resonant driving, $\omega_{d} = \omega$.
We show the evolution of the probability to find the oscillator in the position $x$, Fig. \ref{fig:ResonantDriving}(a); numerical values are shown in shades of green while the analytic expectation value for the position is shown in a solid blue line with thinner lines for its standard deviation. The expectation values of position and momentum from analytic calculations, 
\begin{align}
    \langle \hat{x}(t) \rangle =& -\frac{\Omega t}{2 m \omega} \sin \left( \omega t \right), \\
     \langle \hat{p}(t) \rangle =& -\frac{\Omega}{2 \omega} \left[ \omega t \cos \left( \omega t \right) + \sin \left( \omega t \right) \right], 
\end{align}
are shown in solid blue and red lines, while numerical simulations are shown in blue and red squares, in that order, with its respective variances as shaded regions for analytic and whiskers for numeric values in Fig. \ref{fig:ResonantDriving}(b).
We show the uncertainty principle from analytic and numerical results in solid black line and squares, in that order, in Fig. \ref{fig:ResonantDriving}(c).  

\begin{figure}[ht]
 	\centering
 	\includegraphics
 	{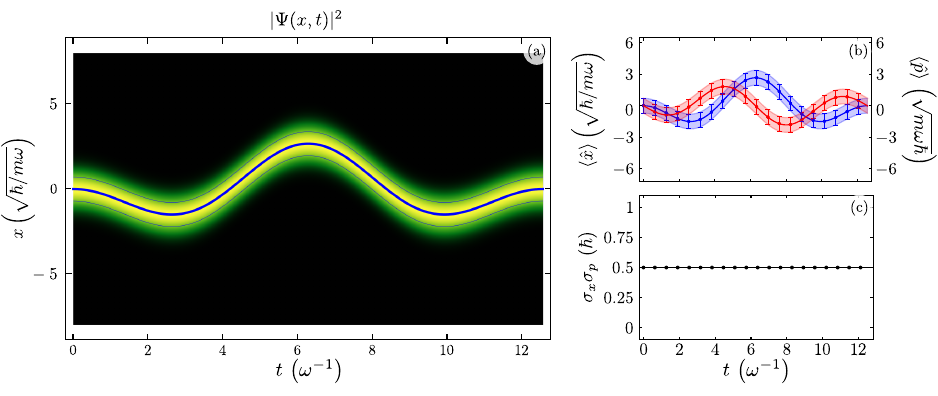}
    \caption{Time evolution of (a) the probability amplitude to find the resonator in the position x $\vert \psi(x) \vert^{2}$, (b) expectation value for the position, $\langle x(t) \rangle$ in blue, and momentum, $\langle p(t) \rangle$ in red, with solid lines and squares for analytic and numeric results, in that order; we show standard deviations, $\sigma_{x}(t)$ and $\sigma_{p}(t)$, as shaded regions or error bars for analytic and numeric results, and (c) Heisenberg uncertainty relation, $\sigma_{x}(t) \sigma_{p}(t)$, for an initial state provided by the ground state of the standard harmonic oscillator for an off-resonant driven QHO, $\omega_{d} = \omega/2$.}\label{fig:OffResonantDriving}
\end{figure}

Figure \ref{fig:OffResonantDriving} shows the evolution of the ground state under far off-resonance driving, $\omega_{d} = \omega / 2$, following the same color coding.
We show the evolution of the probability to find the oscillator in the position $x$, Fig. \ref{fig:OffResonantDriving}(a), the expectation values of position and momentum from analytic and numeric calculations, Fig. \ref{fig:OffResonantDriving}(b), and the uncertainty principle, Fig. \ref{fig:OffResonantDriving}(c).

\begin{figure}[ht]
 	\centering
 	\includegraphics{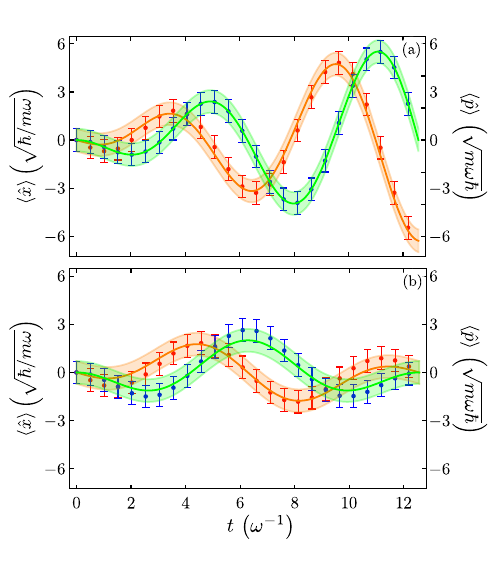}
    \caption{Time evolution of the expected value for our exact analytic solution, box and whisker points, and the analytic rotating wave approximation, solid lines and shadows under (a) resonant driving of the position, $\omega_{d} = \omega$, and  off-resonant driving of the position, $\omega_{d} = \omega/2$. Blue (red) box and whisker points show our exact analytic values and Green (orange) solid lines and shadows the analytic values under the RWA for position (momentum). }\label{fig:ExactvsRWA}
\end{figure}

For the sake of curiosity, we compare the results from our exact solution to the driven quantum harmonic oscillator and the one using the rotating wave approximation (RWA) in Fig. \ref{fig:ExactvsRWA}.
We show our analytic results for the driven QHO without the RWA as box and whisker points and the expectation value of position (momentum) in solid green (orange) lines and its variance in shadow green (orange) under the RWA. 
Driving the position of the QHO under resonance conditions, Fig. \ref{fig:ExactvsRWA}(a) produce no relative error for the expected value of the position but an error in the expected value of the momentum appears soon after the initial time. 
Resonant driving of the momentum produces a similar behavior, the expectation value of the momentum will show no relative error between exact and RWA values.
Off-resonant driving, Fig. \ref{fig:ExactvsRWA}(b), shows relative error for both expected values of position and momentum for driving in the position or momentum.

\subsection{Caldirola--Kanai oscillator}
Now, let us consider the historical Caldirola--Kanai QHO \cite{Caldirola1941p393,Kanai1948p440}, 
\begin{equation}\label{eq:CK_oscillator}
    \hat{H}{\left(t \right)} = \frac{1}{2m e^{\gamma t}}\hat{p}^{2} + \frac{1}{2} m e^{\gamma t} \omega^{2} \hat{x}^{2},
\end{equation}
where the mass is scaled by a time-dependent factor with an exponential form with constant mass $m$ and parameter $\gamma$ with units of $\mathrm{frequency}$ that usually is a decay process, $\gamma <  0$, a constant frequency, $\omega(t)= \omega$ and null values for all other processes, $\alpha_{x}(t) = \alpha_{p}(t) = \alpha_{xp}(t) = \alpha_{0}(t) = 0$.

Before we focus on our approach, we want to show that it is straightforward to realize that an unitary squeezing,
\begin{equation}
    \vert{\psi(t)}\rangle  = e^{\frac{i}{\hbar}\frac{\gamma t}{4}\left\{\hat{x},\hat{p}\right\}}\vert{\phi}(t)\rangle,
\end{equation}
yields an effective time-independent parametric QHO, 
\begin{align}
      \hat{H} = \frac{1}{2m}\hat{p}^{2} + \frac{1}{2}m\omega^{2} \hat{x}^{2} + \frac{\gamma}{4} \left\{ \hat{x}, \hat{p} \right\} ,
\end{align}
feasible of diagonalization using our time-independent approach in the Results Section, 
\begin{equation}
\hat{H}_{5} = \frac{1}{2 m_{5}} \hat{p}^{2} + \frac{1}{2} m_{5} \omega_{5}^{2} \hat{x}^2,
\end{equation}
with effective constant mass and frequency,
\begin{equation}
    \begin{aligned}
        m_{5} &= \frac{m\omega}{\omega + \frac{\gamma}{2} }, \\
        \omega_{5}^{2} &= \omega^{2} - \frac{\gamma^{2}}{4} ,
    \end{aligned}
\end{equation}
to study, for example, the energy spectrum for the model \cite{Colegrave1981p2269}.
This expression allows us to realize that, in order to deal with a well-behaved QHO, the absolute value of the Caldirola-Kanai parameter should be less than twice the QHO resonant frequency, $ \left\vert \gamma \right\vert < 2 \omega$, which agrees with results in the literature \cite{Rath2008p065012,Brown1991p527}.
Furthermore, a generalized Caldirola--Kanai QHO, 
\begin{equation}
H_{CK} = \frac{1}{2 m e^{-\zeta(t)}} \hat{p}^{2} + \frac{1}{2} m e^{-\zeta(t)} \omega_{0}^{2} \hat{x}^{2},    
\end{equation}
with dimensionless parameter $\zeta(t)$,
under a change of reference frame composed by a scaled displacement of the canonical momentum and a squeezing of the canonical pair, 
\begin{equation}
    \vert \psi_{CK} \rangle = e^{- \frac{i}{4 \hbar} \zeta(t) \left\{\hat{x} , \hat{p} \right\}}  e^{ \frac{i}{\hbar} \frac{m\dot{\zeta(t)}}{4}  \hat{x}^{2}} \vert \psi_{PT} \rangle,
\end{equation}
takes the form of the Paul trap Hamiltonian, 
\begin{equation}
H_{PT} = \frac{1}{2 m} p^{2} + \frac{1}{2} m \omega^{2}(t) q^{2},   
\end{equation}
with effective frequency, 
\begin{equation}
\omega^{2}(t) = \omega_{0}^{2} + \frac{ \dot{\zeta}^2(t)}{4} + \frac{\ddot{\zeta}(t)}{2}, \end{equation}
in terms of the time-dependent Caldirola-Kanai parameter as long as it yields a positive real squared effective frequency for all times under the unitary transformation.

We want to test our approach starting from the Caldirola--Kanai QHO and realizing that the absence of driving and energy bias renders the first steps into the identity and allows us to start from the Hamiltonian class in Eq.(\ref{eq:QHOSqueezedFrame}) with null parametric driving and a time-dependent mass with linear exponential time-dependence that yields the following expectation values for the position and momentum, 
\begin{equation}
\begin{aligned}
    \langle \hat{x} (t) \rangle &= \frac{1}{2 e^{\frac{\gamma t}{2}}} \left\{ \sin{\left(\omega_{5}t\right)} \left[ \Gamma_{-} \langle \hat{x} (0) \rangle + \Gamma_{+} \frac{\langle \hat{p} (0) \rangle}{m\omega} \right] + 2 \cos{\left(\omega_{5}t\right)} \langle \hat{x} (0) \rangle \right\}, \\
    \langle \hat{p} (t) \rangle &= \frac{1}{2 e^{\frac{\gamma t}{2}}} \left\{ 2 \cos{\left(\omega_{5}t\right)} \langle \hat{p} (0) \rangle - \sin{\left(\omega_{5}t\right)} \left[ \Gamma_{-} \langle \hat{p} (0) \rangle + \Gamma_{+} m\omega \langle \hat{x} (0) \rangle \right] \right\},
\end{aligned}
\end{equation}
where, for the sake of space, we have defined the auxiliary parameters,
\begin{equation}
    \Gamma_{\pm} = \frac{m\omega}{m_{5}\omega_{5}} \pm \frac{m_{5}\omega_{5}}{m\omega}.
\end{equation}
The variances for the position and momentum,
\begin{equation}
\begin{aligned}
    \sigma_{x}^{2}(t) = \frac{1}{4e^{\gamma t}} &\left\{ \left[ 2 \cos{\left(\omega_{5} t \right)} + \Gamma_{-} \sin{\left(\omega_{5} t\right)} \right]^{2} \sigma_{x}^{2}(0)  + \frac{\Gamma_{+}^{2}}{m^{2}\omega^{2}} \sin^{2}{\left(\omega_{5} t\right)} \sigma_{p}^{2}(0) + \right. \ldots \\
    &\left. \frac{\Gamma_{+}}{m \omega} \left[ \Gamma_{-}\sin^{2}{\left(\omega_{5} t\right)} +\sin{\left( 2\omega_{5} t\right)}\right]\left[ \langle \left\{\hat{x}(0), \hat{p}(0) \right\} \rangle - 2 \langle \hat{x}(0) \rangle \langle  \hat{p}(0) \rangle \right]\right\}, \\
    \sigma_{p}^{2}(t) = \frac{e^{\gamma t}}{4} &\left\{ \left[ 2 \cos{\left(\omega_{5} t \right)} - \Gamma_{-} \sin{\left(\omega_{5} t\right)} \right]^{2} \sigma_{p}^{2}(0) + m^{2}\omega^{2}\Gamma_{+}^{2} \sin^{2}{\left(\omega_{5} t\right)} \sigma_{x}^{2}(0) + \right. \ldots \\
    &\left. m\omega\Gamma_{+} \left[ \Gamma_{-}\sin^{2}{\left(\omega_{5} t\right)} - \sin{\left( 2\omega_{5} t\right)}\right]\left[ \langle \left\{\hat{x}(0), \hat{p}(0) \right\} \rangle - 2 \langle \hat{x}(0) \rangle \langle  \hat{p}(0) \rangle \right]\right\},
\end{aligned}
\end{equation}
allows us to realize that, depending on the initial state, it will be probable to obtain a harmonic oscillation of Heisenberg uncertainty with the effective frequency of the Caldirola-Kanai QHO.

Using the ground state of the standard QHO in Eq.(\ref{eq:groundState}) as the initial state, the expectation values of the position and momentum are zero, 
\begin{equation}
\begin{aligned}
    \langle \hat{x}(t) \rangle &= 0, \\
    \langle \hat{p}(t) \rangle &= 0, \\
\end{aligned}
\end{equation}
and their variances,
\begin{equation}
\begin{aligned}
    \sigma_{x}^{2}(t) &= \frac{\hbar}{8m\omega e^{\gamma t}} \left[   4\cos{\left(2\omega_{5} t \right)} + 2\Gamma_{-} \sin{\left(2\omega_{5} t\right)} + 2 \Gamma_{+}^{2} \sin^{2}{\left(\omega_{5} t\right)} \right], \\
    \sigma_{p}^{2}(t) &= \frac{m\omega\hbar e^{\gamma t}}{8} \left[ 4 \cos{\left(2\omega_{5} t \right)} - 2\Gamma_{-} \sin{\left(2\omega_{5} t\right)} + 2 \Gamma_{+}^{2} \sin^{2}{\left(\omega_{5} t\right)} \right],
\end{aligned}
\end{equation}
will decrease and increase (increase and decrease) exponentially for a negative (positive) Caldirola-Kanai parameter $\gamma$ for the position and momentum, in that order.
They also suggest that the Heisenberg uncertainty relation for the variances,
\begin{equation}
\begin{aligned}
    \sigma_{x}^{2}(t) \sigma_{p}^{2}(t) &= \frac{\hbar^2}{64}\left[4\left(2\cos{\left(2\omega_{5}t\right)} + \Gamma_{+} \sin{\left(\omega_{5}t\right)}\right)^{2}\ - 4\Gamma^{2}_{-} \sin{\left(2\omega_{5}t\right)} \right],
\end{aligned}
\end{equation}
will oscillate harmonically with the effective frequency $\omega_{5}$ for the ground initial state, and will minimize it, $\sigma_{x}^{2}(t) \sigma_{p}^{2}(t) = \hbar^2 / 4$, periodically for $\omega_{5} t = m \pi$. 

\begin{figure}[ht]
    \centering
    \includegraphics{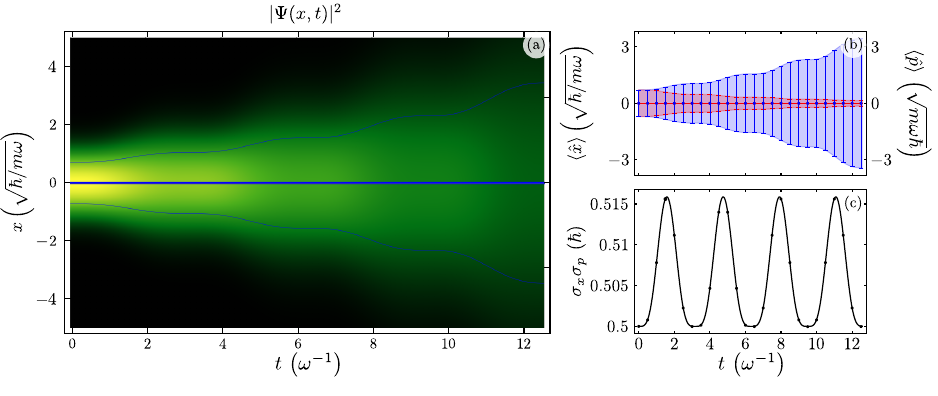}
    \caption{Time evolution of (a) the probability amplitude to find the resonator in the position x $\vert \psi(x) \vert^{2}$, (b) expectation value for the position, $\langle x(t) \rangle$ in blue, and momentum, $\langle p(t) \rangle$ in red, with solid lines and squares for analytic and numeric results, in that order; we show standard deviations, $\sigma_{x}(t)$ and $\sigma_{p}(t)$, as shaded regions or error bars for analytic and numeric results, and (c) Heisenberg uncertainty relation, $\sigma_{x}(t) \sigma_{p}(t)$, for an initial state provided by the ground state of the standard harmonic oscillator in a Caldirola-Kanai QHO with $\gamma = - \omega / 4 $.}
    \label{fig:Ck}
\end{figure}

Figure \ref{fig:Ck} shows the evolution of the ground state of the standard QHO under Caldirola-Kanai QHO dynamics following the color code in Fig. \ref{fig:ResonantDriving} and Fig. \ref{fig:OffResonantDriving}. The probability amplitude is centered around the original, $x=0$, and disperses as it propagates undergoing oscillations, Fig. \ref{fig:Ck}(a).
Figure \ref{fig:Ck}(b) shows the expectation values for the position and momentum from analytic calculations, solid blue and red lines, and numeric simulations, blue and red squares in that order, with their respective variances. 
We show the uncertainty principle from analytic and numeric results, solid black line and squares in that order, that undergoes a single frequency oscillation as expected, in Fig. \ref{fig:Ck}(c).

\subsection{Quadratic time-dependent oscillator }
\begin{figure}[ht]
    \centering
    \includegraphics{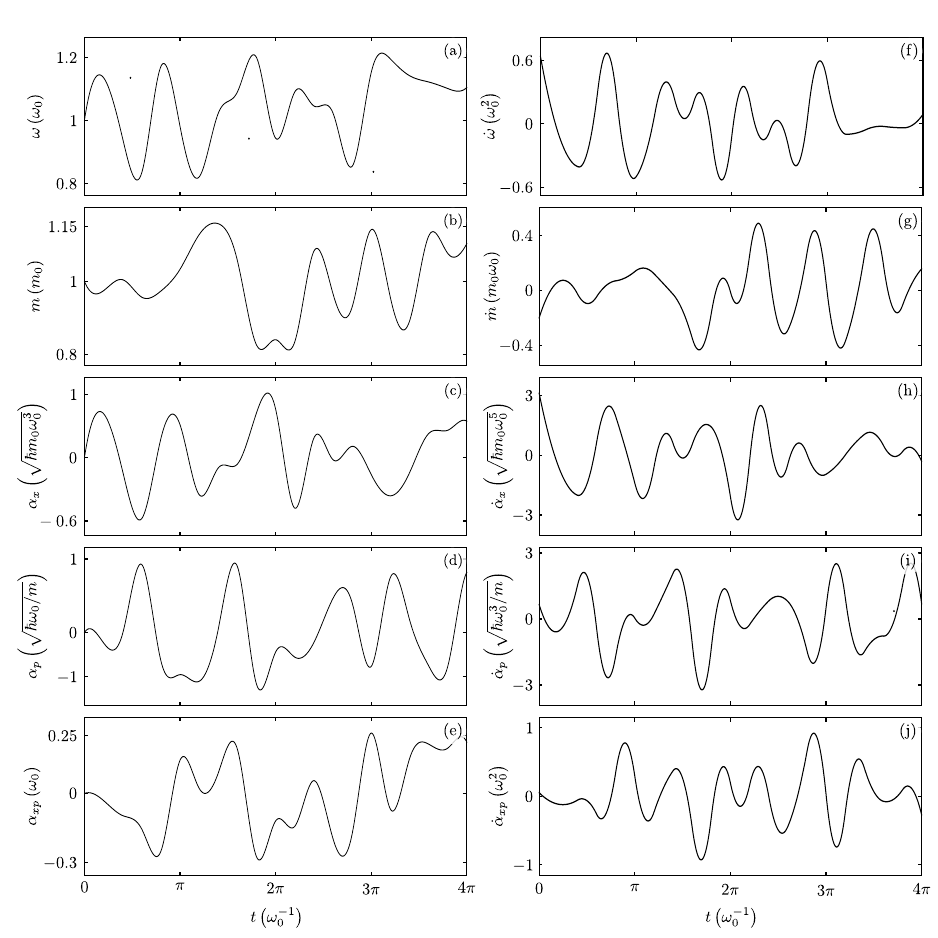}
    \caption{An example of (a)-(e) random time-dependent parameter functions and (f)-(j) its derivatives.}
    \label{fig:Extra1}
\end{figure}
\begin{figure}[ht]
    \centering
    \includegraphics{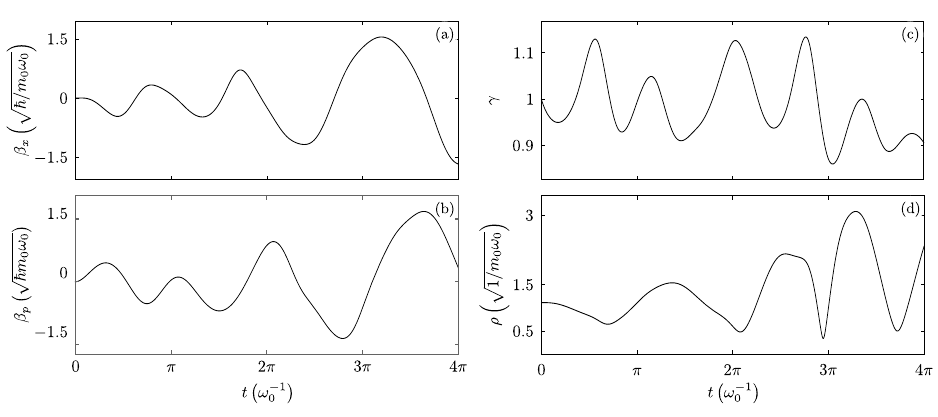}
    \caption{Auxiliary time-dependent functions obtained numerically, (a) position displacement parameter $\beta_{x}(t)$, (b) momentum displacement parameter $\beta_{p}(t)$, (c) squeezing parameter $\gamma(t)$, and (d) Ermakov parameter $\rho(t)$.}
    \label{fig:Extra2}
\end{figure}
\begin{figure}[ht]
    \centering
    \includegraphics{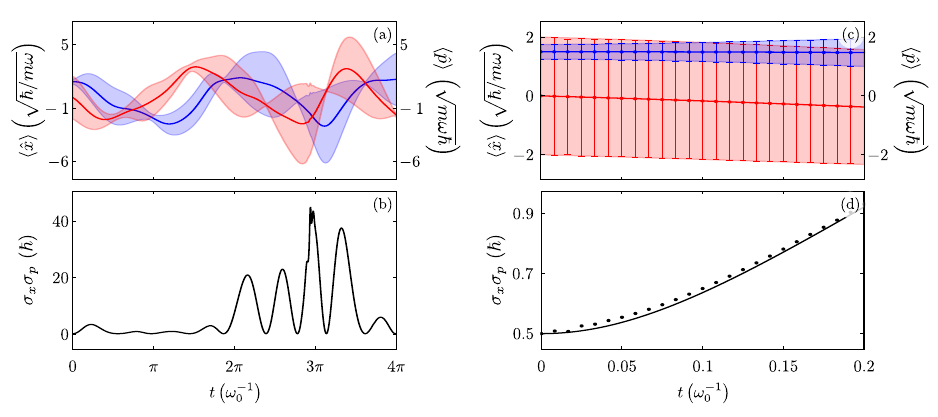}
    \caption{Time evolution of (a) expectation value for the position, $\langle x(t) \rangle$ in blue, and momentum, $\langle p(t) \rangle$ in red, we show standard deviations, $\sigma_{x}(t)$ and $\sigma_{p}(t)$, as shaded regions, and (b) Heisenberg uncertainty relation from a hybrid of analytic and numeric results.
    Time evolution of (c) expectation value for the position, $\langle x(t) \rangle$ in blue, and momentum, $\langle p(t) \rangle$ in red, we show standard deviations, $\sigma_{x}(t)$ and $\sigma_{p}(t)$, as error bars, and (d) Heisenberg uncertainty relation from numeric results in the laboratory frame. The initial state provided is the ground state of the standard harmonic oscillator for a QHO with random time-dependent parameter functions provided by Fig. \ref{fig:Extra1}.}
    \label{fig:Extra3}
\end{figure}

In order to further demonstrate the advantages of our approach, let us explore the full system.
We generate a random sequence of values for each one of the time-dependent parameters, that is, mass, frequency, driving strength, and parametric pumping, and use spline interpolation to generate continuous, differentiable parameter functions for our Hamiltonian class model, Fig. \ref{fig:Extra1}.
We use these as input to numerically generate the auxiliary functions for our analytic factorization scheme, Fig. \ref{fig:Extra2}, and, in consequence, find the observables of interest, Fig. \ref{fig:Extra3}(a) and Fig. \ref{fig:Extra3}(b).
We want to stress that numerically solving Schr\"odinger equation for our model usually yields a stiff problem where the differential equation is unstable under standard numerical methods and we are able to recover expectation values for just small evolution times, Fig. \ref{fig:Extra3}(c) and Fig. \ref{fig:Extra3}(d), while our method produces differential equation sets amenable to solution by standard numerical methods.
These results may be of importance in Quantum technologies that usually involve complex open systems.
There, the system of interest interacts with other systems or the environment, making it challenging to model their behavior accurately.
Machine learning algorithms may be used to make predictions and control their behavior.
However, the amount of experimental data to train them may be limited.
In these circumstances, combining analytic results for closed systems under random perturbations and experimental data from systems weakly coupled to their environment may be useful to generate large and more diverse training data sets which could improve the accuracy of machine learning models.

\section{Conclusions}\label{sec:Conclusion}
We developed a procedure to diagonalize a general quadratic time-dependent QHO using a Lie algebraic approach provided by unitary transformations.
In general, our approach yields a set of differential equations for the parameters in our unitary transformations in terms of the time-dependent parameters of our Hamiltonian model.
Our approach allows for the analytic solution of some particular realizations of our quadratic QHO model and the numeric solution for well-behaved time-dependent functions providing positive definite effective masses and frequencies in each step of the process. 
In addition, our method provides in a straightforward manner the time evolution for the expectation values of observables and their variances. 

We constructed the analytic closed-form solution of two iconic models. 
The periodically driven QHO without the rotating wave approximation and the Caldirola--Kanai QHO. 
Our procedure allows for finding analytic auxiliary functions for both models. 
They help us predict the expectation values for the position and momentum, as well as their variances, to good agreement with numerical solutions in any given parameter regime. 
In addition, we are able to show that there exists a class of Caldirola--Kanai QHO that takes the form of the Paul trap Hamiltonian under a suitable change of reference frame provided by a unitary transformation.
We also show that a hybrid version of our approach, that is, our analytical results combined with numerical methods to determine the auxiliary functions, is able to provide expectation values for the variables of interest for a random realization of our general quadratic time-dependent QHO with well-behaved parameters that yields an Schr\"odinger equation that is unstable for standard numerical methods.

We hope our results help with the analysis of driven parametric systems that appear, for example, in circuit quantum electrodynamics. 
In these systems, the rotating wave approximation is a common source of predictive errors beyond the regime of near-resonance and small driving strength.
Furthermore, the hybrid analytic--numeric version of our approach may be useful for generating training data sets for expert systems used, for example, to control the dynamics of experiments related to quantum technologies.
\begin{acknowledgments}
F.~E.~O. acknowledges financial support from CONACyT and thanks U.N.N for study leave support.
E.~G.~H, J.~A.~R.-G., G.~J.~R, and E.~R.~N acknowledge support from the Bachelor Research Internship program at Tecnológico de Monterrey under the project  “Symmetry-based design of classical and quantum optical devices".
\end{acknowledgments}


%


\end{document}